\documentclass{aastex}
\usepackage{spr-astr-addons}
\usepackage{graphicx}
\usepackage{txfonts}
\usepackage{xcolor}
\usepackage{threeparttable}
\usepackage{tikz}
\usepackage{rotating}
\usepackage{lscape}
\usepackage{adjustbox}
\usepackage{graphicx}
\usepackage[version=4]{mhchem}
\usepackage[normalem]{ulem}
\usepackage{float}
\usepackage{natbib}
\pdfoutput=1 
\usepackage{amsmath,amstext,amssymb}
\usepackage[T1]{fontenc}
\usepackage[colorlinks=true, allcolors=blue]{hyperref}

\RequirePackage{color}

\begin{document}

\title{Timing and spectral studies of the X-ray pulsar 2S 1417--624 during the outburst in 2021}
	\shorttitle{2S 1417--624 during the outburst in 2021}
	\shortauthors{Mandal \& Pal}

	\author{Manoj Mandal\altaffilmark{1}} \and \author{Sabyasachi Pal\altaffilmark{1}}

	\altaffiltext{1}{Midnapore City College, Kuturia, Bhadutala, Paschim Medinipur, West Bengal, 721129, India \\Email: {manojmandal@mcconline.org.in (MM)}, {sabya.pal@gmail.com (SP)}}

\abstract
{We study the timing and spectral properties of the X-ray pulsar
2S 1417--624 during the recent outburst in January 2021 based on the Neutron
Star Interior Composition Explorer (NICER) observation. We also used some
early data from the 2018 outburst to compare different temporal and spectral
properties. The evolution of the spin period and pulsed flux is studied
with \textit{Fermi}/GBM during the outburst and the spin-up rate is found
to be varied between $\simeq(0.8\hbox{--}1.8)\times 10^{-11}\text{ Hz}\,\text{s}^{-1}$. The pulse profile shows energy dependence and variability. The pulse profile shows multiple peaks and dips which evolve with energy. The evolution of the spectral state of this source is also studied using the hardness intensity
diagram (HID). The HID shows a transition from the horizontal to the diagonal
branch, which implies the source went through a state transition from the
subcritical to supercritical accretion regime. The \textit{NICER} energy spectrum
is well described by a composite model of a power-law with a higher cut-off
energy and blackbody components along with a photo-electric absorption
component. An iron emission line is detected near 6.4~keV in the \textit{NICER}
spectrum with an equivalent width of $\sim $0.05~keV. The photon index
shows an anti-correlation with flux below the critical flux. The mass accretion
rate is estimated to be $\simeq 1.3 \times  10^{17}\text{ g}\,\text{s}^{-1}$ near
the peak of the outburst. We have found a positive correlation between
the pulse frequency derivatives and luminosity. The Ghosh and Lamb model
is applied to estimate the magnetic field at different spin-up rates, which
is compared to the earlier estimated magnetic field at a relatively high
mass accretion rate. The magnetic field is estimated to be $\simeq 10^{14}$~G from the torque-luminosity model using the distance estimated by Gaia,
which is comparatively higher than most of the other Be/XBPs.}




\maketitle

\section{Introduction}
\label{intro}
The X-ray transient pulsar 2S 1417--624 was discovered using the Small
Astronomy Satellite \textit{(SAS--3)} in 1978 \citep{Ap80}. Several outbursts
from the source were observed by the Burst and Transient Source Experiment
(\texttt{BATSE}) and the Rossi X-ray Timing Explorer (\textit{RXTE})
\citep{Fi96a, Gu18}. Earlier, an X-ray pulsation at $\sim $17.5~s was detected
from the source light curve with an orbital period of $\sim $42 days
\citep{Fi96a}. The orbital parameters of the binary system were improved
by \citet{Ra10} using \textit{RXTE} during the giant outburst in 1999. The
source is located at a distance of $\sim $9.9~kpc provided by \textit{Gaia}
\citep{Ba18}. The accretion-powered X-ray pulsar 2S 1417--624 went through
a giant outburst in 2009, and different timing and spectral properties
were studied using \textit{RXTE} \citep{Gu18}. During this outburst, the pulse
profile showed energy and luminosity dependence, and the pulse profile
evolved from a double-peak feature at lower luminosity to a triple-peak
feature at higher luminosity and back to a double-peak feature during the
decay phase of the outburst. The variation of pulse fraction was studied
with flux, which showed an anti-correlation with source flux during the
outburst.

During the \textit{MAXI} observation, the strong energy dependence of the
pulse profile was observed, and the four-peaked pulse profile at lower
energies evolved into a double peak feature at higher energies. The pulse
fraction showed an anti-correlation with luminosity, which was similar
to the previous giant outburst in 2009 \citep{Gu19}. Variability in the
pulse profile was also observed from the \textit{NICER} observations, and
the pulse profile evolved significantly with luminosity and energy
\citep{Ji20}. The magnetic field was estimated to be $\sim 7\times 10^{12}$~G for a source distance of $\sim $20 kpc by considering the spin-up due
to the accretion torque.

The critical luminosity ($L_{\mathrm{crit}}$) of a source is crucial in
defining two accretion regimes. The source luminosity is lower than the
critical luminosity in the subcritical regime, and at the critical luminosity,
a state transition from the subcritical to the supercritical regime occurs.
Near the critical luminosity, the pulse profile, pulsed fraction, and beaming
patterns change significantly. The state transition can be probed using
the hardness intensity diagram (HID). During the state transition, the
HID shows a transition from the horizontal branch (low luminosity state)
to a diagonal branch (high luminosity state), which was observed earlier
for several sources \citep{Re13}.

During the 2018 giant outburst, 2S 1417--624 was studied using \textit{Swift},
\textit{MAXI} \citep{Gu19}, \textit{NICER} \citep{Se22}, and \textit{Insight-HXMT}
\citep{Ji20}. \citet{Se22} reported a state transition from a subcritical
to a supercritical regime during the 2018 outburst. A significant evolution
of different spectral parameters was found near the critical X-ray flux
(unabsorbed) of $\sim $0.7$\times 10^{-9}$~erg\,cm$^{-2}$\,s$^{-1}$ using
\textit{NICER} observations (0.8--12~keV).

Recently, the source went through an outburst in 2021 that was the strongest
after the giant outburst of 2018. The outburst was detected by \textit{Fermi}/GBM,
Burst Alert Telescope (BAT) onboard \textit{Swift}, and Gas Slit Camera (GSC)
onboard \textit{MAXI} on January 2021 \citep{Ha21}. The X-ray flux started
to increase from the early last week of January 2021, and the duration
of the outburst was nearly three months. In this paper, we study the timing
and spectral properties of the pulsar 2S 1417--624 during the recent outburst
in 2021 using \textit{NICER} observations and compare different timing and
spectral properties with the giant outburst of 2018. We describe the data
reduction and analysis method in Sect.~\ref{sec:obs}. We have presented
the results of the current study in Sect.~\ref{res}. The discussion and
conclusion are summarized in Sect.~\ref{dis} and \ref{con} respectively.

\begin{figure}[t]
\centering{
\includegraphics[width=8cm]{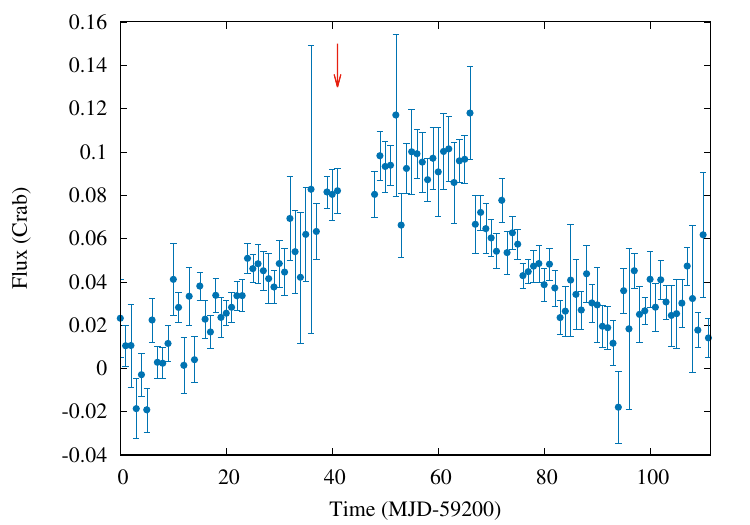}
	\caption{An outburst is detected  from X-ray pulsar 2S 1417--624 by {\it Swift}/BAT (15--50 keV) during January-March, 2021. The red solid arrow shows the time of {\it NICER} observation.}
\label{fig:BAT}}
\end{figure}

\begin{table*}
\centering
	\caption{Log of the \textit{NICER} observations. Observation 1 is the \textit{NICER} observation during the 2021 outburst and the rest of the observations are during the 2018 giant outburst. Flux is estimated in the energy range of 0.8--12~keV in the unit of $10^{-10}$~erg\,cm$^{-2}$\,s$^{-1}$.}
\label{tab:log_table}
\scalebox{0.9}{
\begin{tabular}{ccccc} 

	\hline
	 Start time       & {\it NICER} flux & Exposure  & Obs. ID & Pulsed fraction \\
		      (MJD)          & & (ks) &      &    (\%)          \\
\hline
	 59241.50 & 6.539$\pm$0.004 & 7.165 & 3200130112 (Obs 1) & 27.12$\pm$1.7 \\
 58308.25 & 7.398$\pm$0.006 & 1.745 & 1200130165 (Obs 2) & 35.14$\pm$2.0 \\
	 58296.28 & 9.466$\pm$0.005 & 1.423 & 1200130155 (Obs 3) & 28.64$\pm$1.8\\
	58310.32 & 6.467$\pm$0.009 & 0.632 & 1200130166 (Obs 4) & 32.40$\pm$2.0\\
	58312.44 & 5.935$\pm$0.009 & 0.936 & 1200130168 (Obs 5) & 37.01$\pm$2.1\\
	58317.70 &4.560$\pm$0.020 & 0.959 & 1200130169 (Obs 6) & 43.90$\pm$2.2\\
	58214.77 & 6.844$\pm$0.010 & 0.998 & 1200130104 (Obs 7) & 31.15$\pm$1.8\\
    58274.33 & 11.63$\pm$0.90 & 1.146 & 1200130143 (Obs 8) & 27.86$\pm$2.0\\
    58275.04 & 11.55$\pm$0.90 & 1.646  & 1200130144 (Obs 9) & 20.05$\pm$1.7\\
    58326.12 & 3.95$\pm$0.96 & 1.360 & 1200130175 (Obs 10) & 37.18$\pm$2.0\\
    58328.82 & 3.90$\pm$0.90 & 0.856 & 1200130177 (Obs 11) & 41.93$\pm$2.2\\
\hline
\end{tabular}}
\end{table*}
  
\section{Observation and data analysis}
\label{sec:obs}
We detected an outburst from the X-ray pulsar 2S 1417--624 and followed
the evolution of the outburst using different instruments. We use data
from all-sky X-ray monitors like \textit{Swift}/BAT (15--50~keV), \textit{MAXI}/GSC
(2--20~keV), and \textit{Fermi}/GBM (12--50~keV). We analyzed the \textit{NICER}
data during the rising phase of the 2021 outburst near the peak. We used
the \texttt{HEASOFT} v6.27.2 for the data reduction and analysis. BAT onboard
the \textit{Swift} observatory \citep{Ge04} is sensitive in hard X-ray (15--50~keV) \citep{Kr13}. We have used the results of the BAT transient monitor
during the outburst, which were provided by the BAT team. BAT flux reached
a maximum of $\sim $0.1 Crab during the first week of February 2021.

We have made use of \textit{MAXI}/GSC (2--20~keV) light curves data
\citep{Ma09} to follow up on the outburst and to study the evolution of
spectral states. \textit{MAXI} in-orbit operation was started in 2009, and
nearly 300 pre-registered sources have been monitored at regular intervals
in different energy bands (2--4~keV, 4--10~keV, and 10--20~keV bands).
The data provided by \textit{MAXI}/GSC is averaged for every day. We have
studied the evolution of the hardness of the X-ray pulsar using data from
different energy bands.

The Neutron Star Interior Composition Explorer\break  (\textit{NICER}) was launched
in 2017 and is currently working as an external payload on the International
Space Station. \textit{NICER} consists of one instrument, the X-ray Timing
Instrument (XTI), operating in the soft X-ray region (0.2--12~keV)
\citep{Ge16}. A follow-up observation of 2S 1417--624 was conducted by
\textit{NICER} on January 27, 2021, during the rising phase of the outburst,
close to the peak of the outburst. The details of the \textit{NICER} observation
are tabulated in Table~\ref{tab:log_table}. The processing of raw data
has been done using the \textit{NICERDAS} in \texttt{HEASOFT} v6.27.2. The \textit{NICER}
data are reduced with Calibration Database (CALDB) version xti20200722.
We have created clean event files by applying the standard calibration
and filtering tool \texttt{nicerl2} to the unfiltered data. We have extracted
light curves and spectra using \texttt{XSELECT} from the barycenter corrected
reprocessed clean event file. For the timing analysis, we selected good
time intervals according to the following conditions: ISS not in the South
Atlantic Anomaly (SAA) region, source elevation $>$20$^{\circ}$ above the
Earth limb, source direction at least 30$^{\circ}$ from the bright Earth.

For timing analysis, we have applied barycentric corrections to those events
using the task \texttt{barycorr}. The ancillary response file and response
matrix file of version 20200722 were considered in our spectral analysis.
The background corresponding to each epoch of the observation was simulated
by using the \texttt{nibackgen3C50}\footnote{\url{https://heasarc.gsfc.nasa.gov/docs/nicer/tools/nicer\_bkg_est\_tools.html}}
tool \citep{Re22}. Ancillary response files and response matrix files of
version 20200722 are considered in our spectral analysis. We have used
the latest response files (nixtiref20170601v002.rmf,\break nixtiaveonaxis20170601v004.arf)
for the spectral analysis.

The \textit{Fermi} Gamma-ray Space Telescope operates within a wide energy
range between 8~keV--40~MeV. The Large Area Telescope (LAT) and Gamma-ray
Burst Monitor (GBM) are the two main instruments onboard the \textit{Fermi}
Gamma-ray Space Telescope \citep{Me09}. The GBM is made up of 14 detectors:
12 detectors of Sodium Iodide (NaI) and 2 detectors of Bismuth Germanate
(BGO). In the current study, we have used the spin frequency, frequency
derivative, and 12--50~keV pulsed flux measurements with the \textit{Fermi}/GBM
\citep{Fi09}. The outburst from 2S 1417--624 was also detected with \textit{Fermi}/GBM
from January 2021 and continued for nearly two months with a maximum pulsed
flux of $\sim $0.27~keV\,cm$^{-2}$\,s$^{-1}$ on MJD 59260 as provided by
\textit{Fermi}/GBM \citep{Me09}.

The spin-frequencies are also used, which are provided by the \textit{Fermi}/GBM
team. There were a total of 23 spin frequency ($\nu $) measurements conducted
during our study, and we used 18 measurements, which were at around 3-day
equal intervals, and we have not included the first and last few measurements.
We used a linear function to fit each of the three consecutive frequency
measurements with time. The spin-up rate was calculated from the slope
of the linear function during a 9-day interval, as each pulse frequency
measurement was collected every 3-day interval, using the
$\chi ^{2}$ minimization technique. We repeated this process for the next
three frequency measurements and so on (viz. \citet{Ka20}). Therefore,
we had 6 spin-up rates from 18 spin frequencies.

We have used the average value of total flux for three consecutive points
over the same intervals, which are used to determine $\dot{\nu}$. Finally,
the luminosity is estimated from the X-ray flux for a distance of
$\sim $9.9 kpc. X-ray luminosity of the source is calculated from the count
rate history provided by \textit{Swift}/BAT team \citep{Kr13} by multiplying
a flux conversion factor of $1.13\times 10^{-7}$~erg\,cts$^{-1}$
\citep{Ji20}.

\begin{figure*}
\centering{
\includegraphics[width=6.0cm, angle=270]{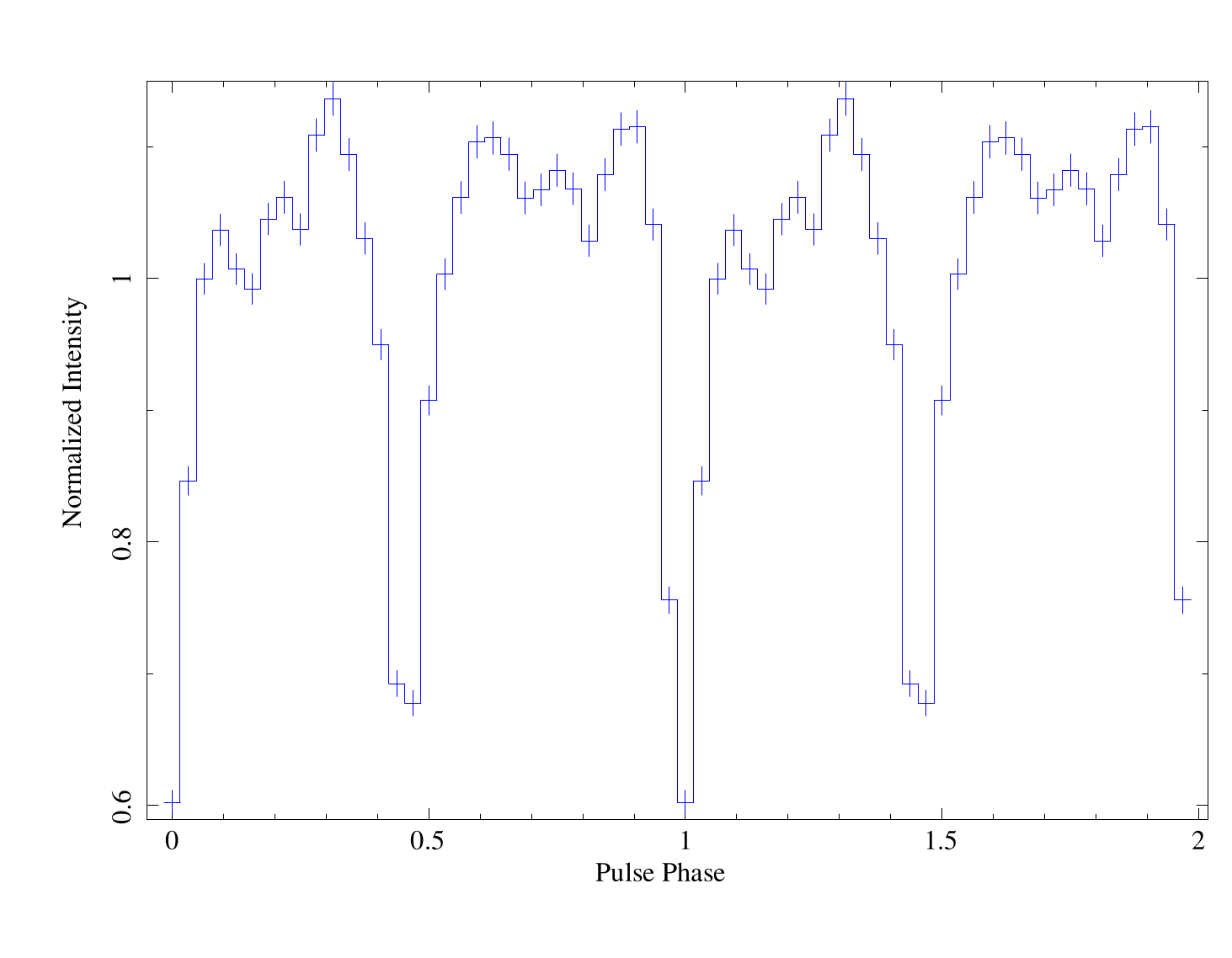}
\includegraphics[width=6.0cm, angle=270]{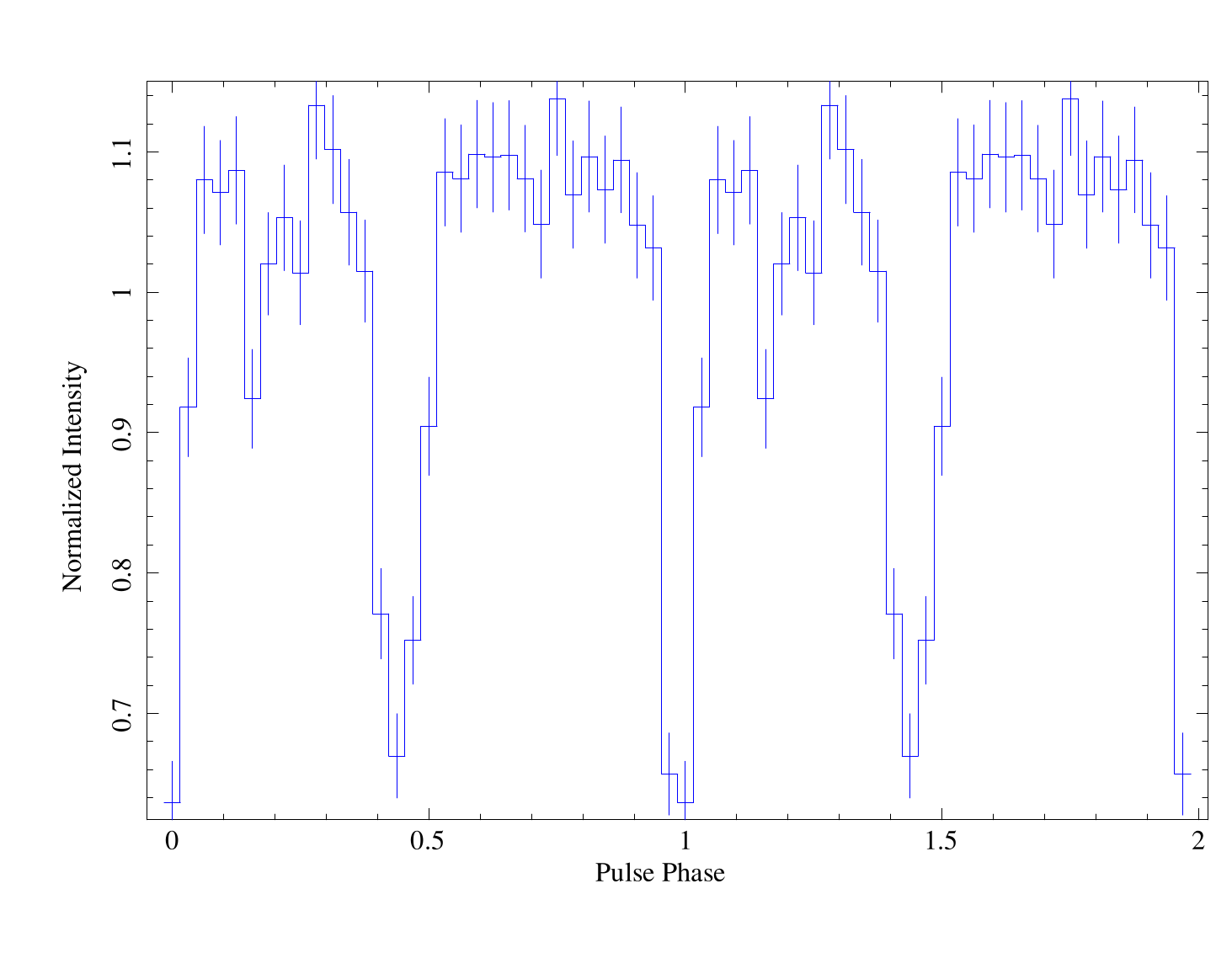}
	\caption{The pulse profile of 2S 1417--624 using \textit{NICER}/XTI. The left
side image shows a pulse profile (0.4--10~keV) during the 2021 outburst
(Obs. ID--3200130112) and the right side image shows a pulse profile (0.4--10~keV) during the 2018 outburst (Obs. ID--1200130104) at a comparable flux
level.}
\label{fig:pulse-profile}}
\end{figure*}

 \begin{figure*}[t]
\centering
\includegraphics[width=6.0cm,angle=270]{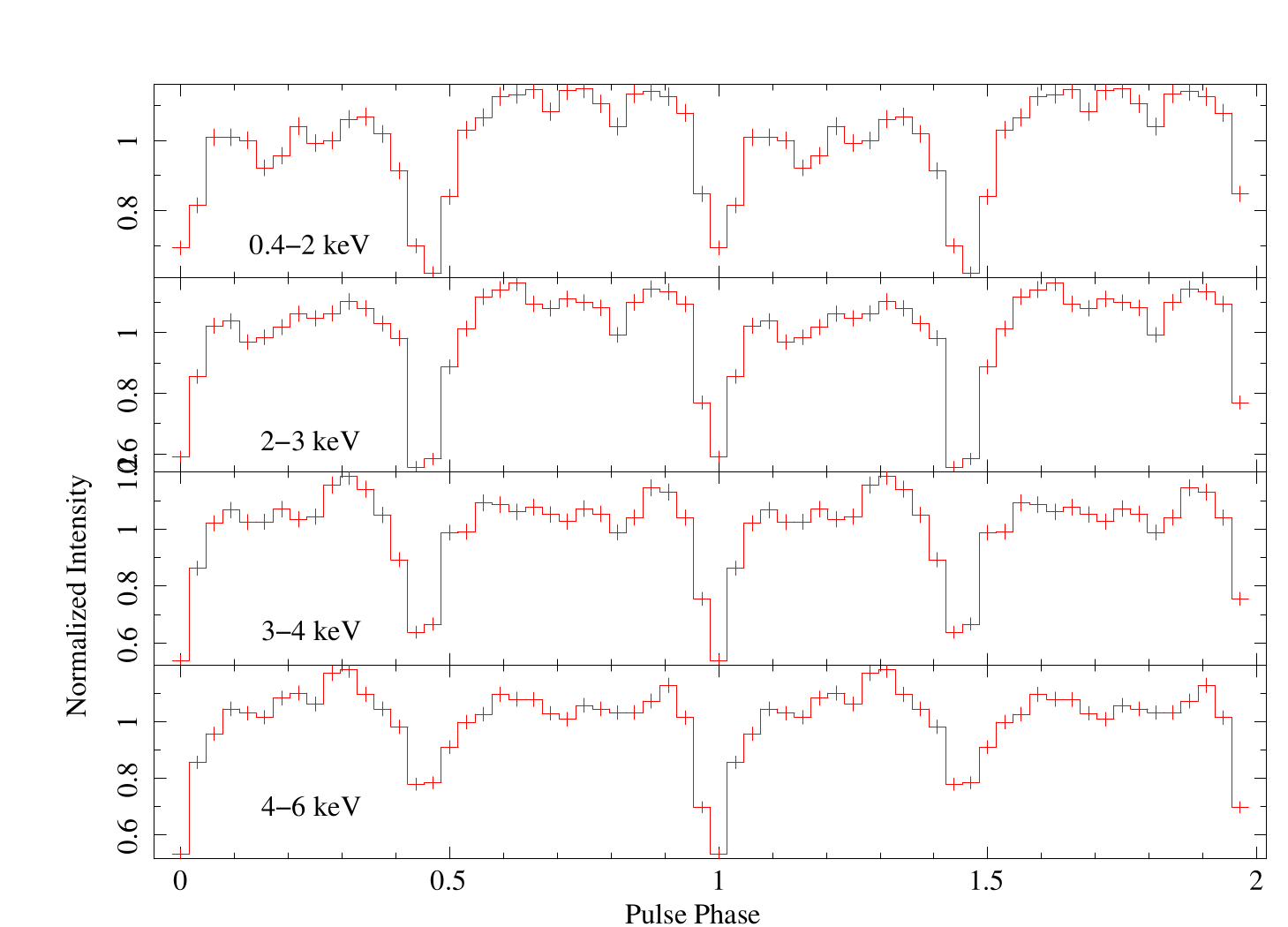}
\includegraphics[width=6.0cm,angle=270]{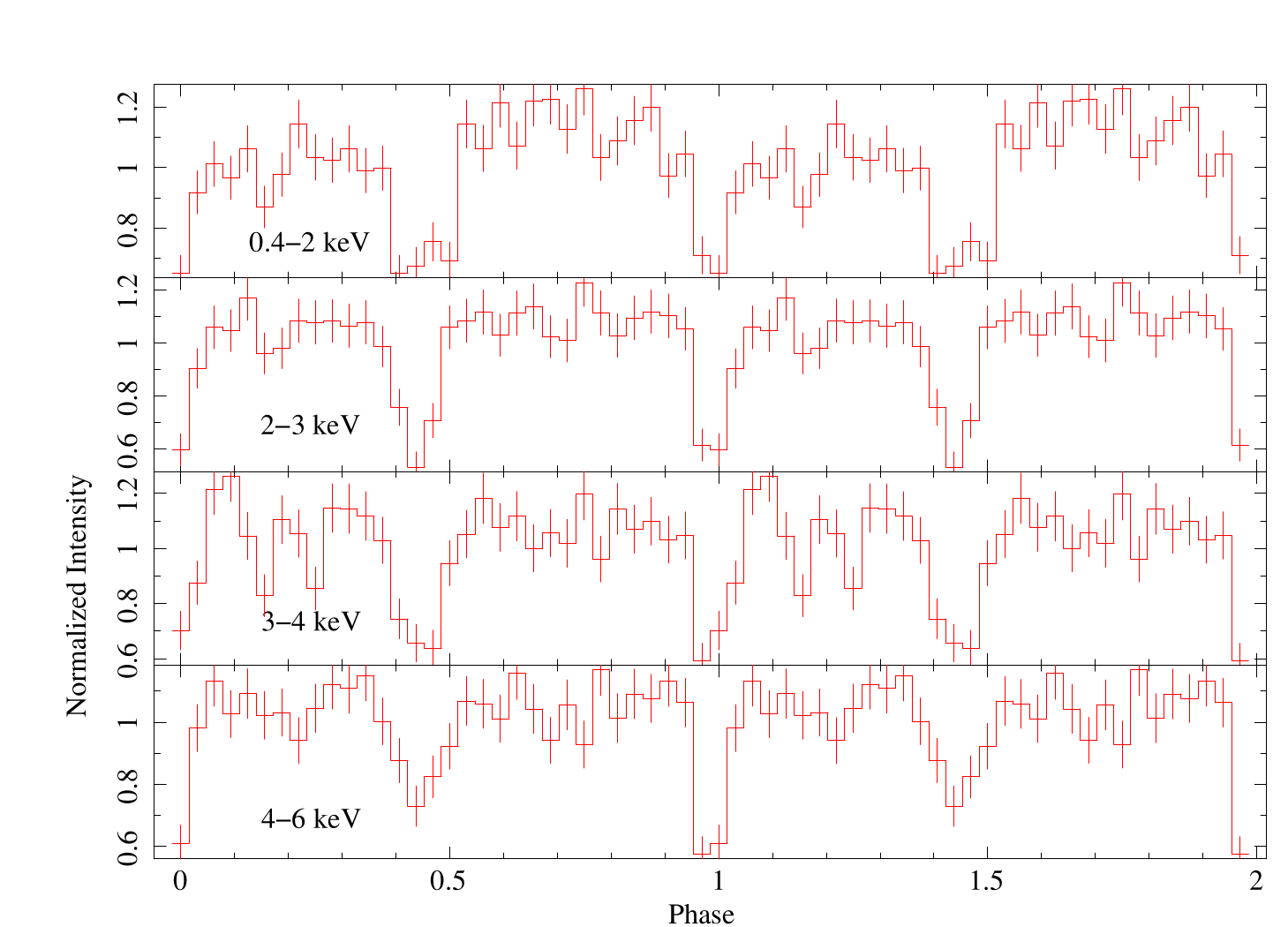}
 \caption{Energy-dependent pulse profiles of the X-ray pulsar 2S 1417--624
using \textit{NICER}/XTI observations. The left-hand side figure shows the
energy-dependent profile during the 2021 outburst and the right-hand side
the figure shows the energy-dependent profile during the 2018 outburst (\textit{NICER}
Obs. 7).}
\label{fig:energy_dependent_pulse_profile}
\end{figure*}

\section{Results}
\label{res}
The X-ray pulsar 2S 1417--624 went through an outburst during January-March
2021, detected by \textit{Fermi}/GBM, \textit{Swift}/BAT,\footnote{\url{https://swift.gsfc.nasa.gov/results/transients/}}
\textit{MAXI}/GSC, which reached a maximum flux during the second week of
February 2021. Figure~\ref{fig:BAT} shows the variation of hard X-ray flux
during the outburst using \textit{Swift}/BAT (15--50~keV). The total duration
of the outburst was around 3 months, which started in early January 2021
and continued till March 2021. We have summarized the results of the timing
and spectral analysis of 2S 1417--624 during the recent outburst in 2021.
We have used \textit{MAXI} \citep{Ma09} final data products (light curves)
as well as the \textit{Fermi} \citep{Fi09, Me09} pulse frequencies and pulsed
flux evolution data for this source.

\subsection{Variation of pulse profile and pulsed fraction}
The light curves were produced using the science event data in different
energy ranges with a bin size of 0.1~s from \textit{NICER} data. We used the
\texttt{efsearch} task in \texttt{FTOOLS} to check for the periodicity in the
time series of the barycenter and background corrected data sets. We used
the\vadjust{\eject} folding method of the light curve over a trial period to get the best
period by $\chi{^{2}}$ maximizing process \citep{Le87} over 32 phase bins
in each period. After getting the best spin period, pulse profiles were
generated using the \texttt{efold} task in \texttt{FTOOLS} by folding light curves
with the best spin period. Uncertainty in the estimated spin period was
computed using the task \texttt{efsearch} in \texttt{FTOOLS} from the chi-square
versus spin period plot \citep{Ra10}. The evolution of the pulse period
and pulsed flux during the outburst was studied using the \textit{Fermi}/GBM.

We have studied the variation of different timing properties of the X-ray
pulsar 2S 1417--624 during the outburst using \textit{NICER} observations.
The spin period of the pulsar during the outburst is found to be $P =
17.3649\pm 0.0001\text{ s}$ using \textit{NICER} data, which is comparable with
the pulse period recorded with \textit{Fermi}/GBM\footnote{\url{https://gammaray.nsstc.nasa.gov/gbm/science/pulsars}}
during the outburst. \textit{Fermi}/GBM found that the period decreased slowly
with the time of the outburst. Figure~\ref{fig:pulse-profile} shows the pulse
profile using \textit{NICER} data in the energy range of 0.4--10~keV, which
consists of multiple broad peaks and narrow dips. We have compared pulse
profiles with the 2018 outburst at the comparable flux level. The left
side of Fig.~\ref{fig:pulse-profile} shows the pulse profile during the
2021 outburst, and the right side of Fig.~\ref{fig:pulse-profile} shows
the pulse profile during the 2018 outburst.

We have looked at the energy dependence of the pulse profiles as well as
the temporal variation of the pulse profile during the outburst. Figure~\ref{fig:energy_dependent_pulse_profile} represents the energy-dependent
pulse profile for four different energy ranges. The variation of the pulse
profile over four energy bands: 0.4--2~keV, 2--3~keV, 3--4~keV, and 4--6~keV is shown in Fig.~\ref{fig:energy_dependent_pulse_profile}. We have
also estimated the pulse profile in the 6--10~keV band, but due to the
low count rate in this band, we have not included this. The pulse profile
shows two broad peaks and dips, which varied with energy. The pulse profile
of the first row (0.4--2~keV) of Fig.~\ref{fig:energy_dependent_pulse_profile} shows two clear dips and two broad
peaks. We compare the energy-resolved pulse profile with the previous giant
outburst of 2018 at comparable flux levels, which also showed two broad
peaks and dips which evolved slightly with energy.

For estimating the pulsed fraction, we used this formula:
\begin{equation}
PF(\%)=
\frac {I_{\mathrm{max}}-I_{\mathrm{min}}} {I_{\mathrm{max}}+I_{\mathrm{min}}}
\times 100
\end{equation}
where $I_{\mathrm{max}}$ and $I_{\mathrm{min}}$ are the maximum and minimum
intensities respectively in the folded light curve.

\begin{figure}[t]
\centering{
\includegraphics[width=8.0cm]{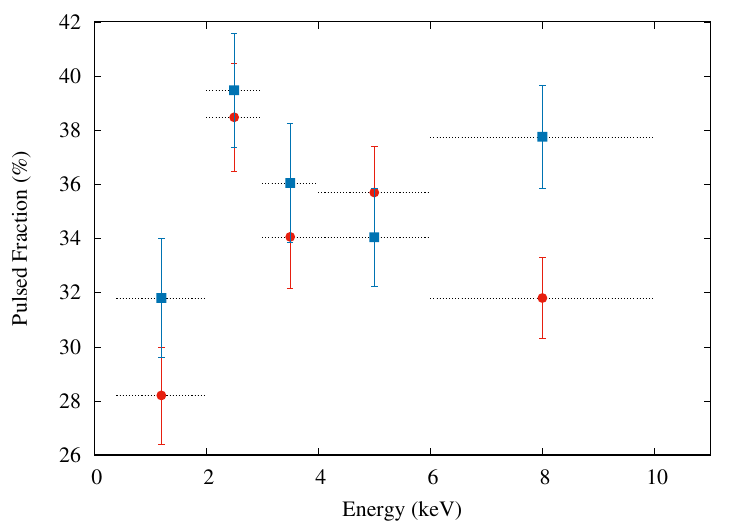}
\caption{Variation of pulsed fraction with energy using \textit{NICER}. The
horizontal lines in all points represent the corresponding energy ranges
for which the pulsed fraction is estimated, and the vertical lines represent
the corresponding error bars of the pulsed fraction. The blue points represent
the PF for the 2018 outburst (Obs. 7, flux $=$ $6.844\times 10^{-10}$~erg\,cm$^{-2}$\,s$^{-1}$ in the energy range 0.8--12~keV) and the red circles represent the PF for the 2021 outburst (Obs. 1, flux $=$ $6.539\times 10^{-10}$~erg\,cm$^{-2}$\,s$^{-1}$ in the energy range 0.8--12~keV).}
\label{fig:pulse_fraction}}
\end{figure}

Figure~\ref{fig:pulse_fraction} shows the variation of pulsed fraction for
different energy ranges for which the energy-resolved pulse profile is
studied. The horizontal bars represent the energy ranges for which the
PF is calculated, and the vertical bars indicate the corresponding error
in measurements. We have compared the value of pulse fraction at the same
energy ranges with the 2018 and 2021 outbursts.

\begin{figure}[t]
\centering{
\includegraphics[width=8.0cm]{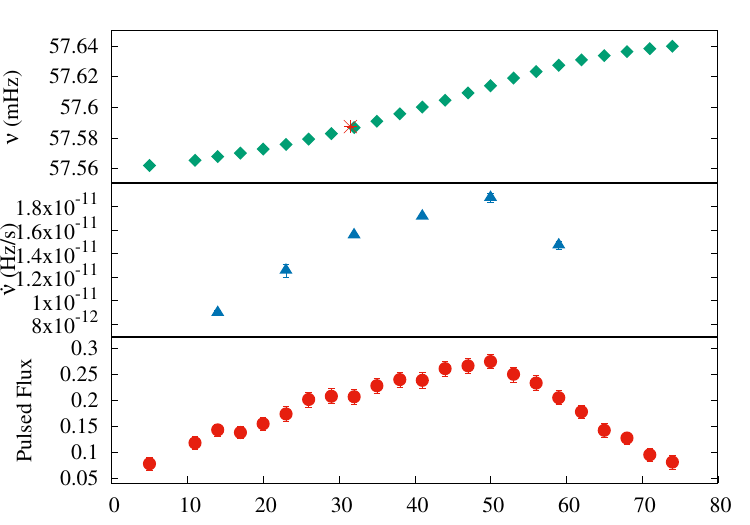}
\caption {The top panel shows the variation of pulse frequency ($\nu $)
of 2S 1417--624 during the outburst using \textit{Fermi}/GBM. The pulse frequency
estimated using \textit{NICER} is shown with a red asterisk. The middle panel
represents the variation of frequency derivatives ($\dot{\nu}$), which
is estimated from \textit{Fermi}/GBM spin period evolution history. The bottom
panel shows the evolution of pulsed flux (12--50~keV) in the unit keV\,cm$^{-2}$\,s$^{-1}$ using \textit{Fermi}/GBM. The vertical bars represent errors in corresponding
measurements.}
\label{fig:freq}}
\end{figure}

We have estimated the pulsed fraction for a few earlier \textit{NICER} observations
during the 2018 outburst. Table~\ref{tab:log_table} (5th column) summarizes
the value of the pulsed fraction. The pulsed fraction shows a trend to
decrease with an increase in flux. The first row of the table corresponds
to the 2021 outburst and the rest of the rows are during the 2018 outburst.
The right-hand side image of Fig.~\ref{fig:PIflux} shows the variation
of pulsed fraction with flux. This indicates that the pulsed fraction is
decreased with an increase in X-ray flux. Figure~\ref{fig:freq} shows the
evolution of the spin frequency and pulsed flux (12--50~keV) during the
outburst of 2021 using Fermi/GBM, which implies that the pulse period of
the X-ray pulsar has decreased slowly with time. We estimated the spin
frequency derivative (as described in Sect.~\ref{sec:obs}), which varied
between $\sim $0.8--$1.8\times 10^{-11}\text{ Hz}\,\text{s}^{-1}$ during the outburst
(shown in the middle panel of Fig.~\ref{fig:freq}).

\begin{figure}[t]
\centering{
\includegraphics[width=7cm]{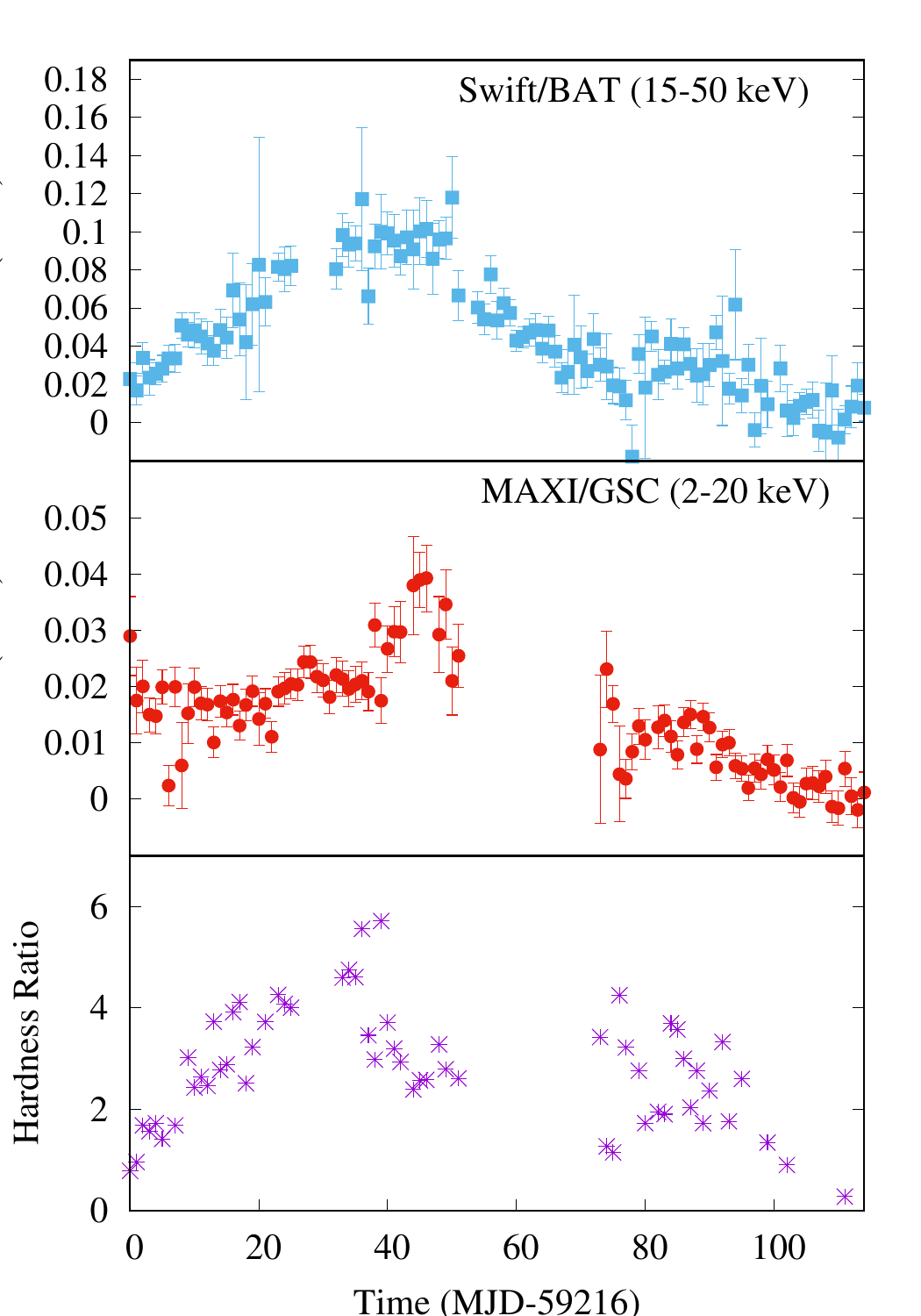}
	\caption{The top row shows the evolution of flux using \textit{Swift}/BAT and
the middle row represents the evolution of flux using \textit{MAXI}/GSC during
the outburst of 2021. The bottom row shows the variation in the hardness
ratio (BAT/MAXI) during the outburst.}
	\label{fig:HR}}
\end{figure}

We have studied the evolution of the spectral state of the source during
the outburst. Figure~\ref{fig:HR} shows the variation of flux using \textit{Swift}/BAT
and \textit{MAXI}/GSC. The top panel of Fig.~\ref{fig:HR} shows the variation
of hard X-ray flux using BAT (15--50~keV), which indicates that the flux
reached a value of $\sim $0.12~crab near the peak of the outburst. The
middle panel of Fig.~\ref{fig:HR} shows the variation of flux using \textit{MAXI}
(2--20~keV), which indicates that the highest flux was $\sim $0.04~crab
near the peak of the outburst. The bottom panel of Fig.~\ref{fig:HR} shows
that the HR varied between 0.1--6 during the outburst.

\begin{figure}[t]
\centering{
\includegraphics[width=8cm]{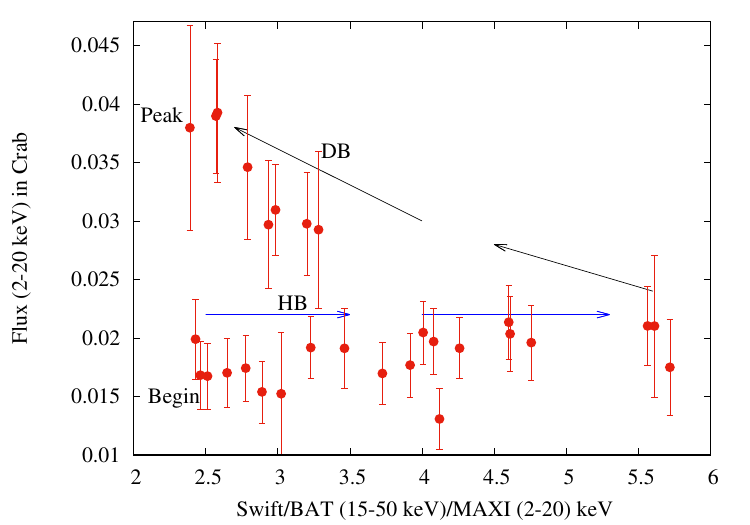}
	\caption{Hardness intensity diagram using \textit{BAT} and \textit{MAXI} flux.
The hardness ratio (\textit{BAT}/\textit{MAXI}) is estimated during the time
MJD 59225--59266 of the outburst. The low luminosity states are represented
by the horizontal branch and the high luminosity states are shown by the
diagonal branch. A transition from a horizontal to a diagonal branch is
visible for this source.}
	\label{fig:HID}}
\end{figure}

The hardness ratio shows a significant variation during the outburst. The
HR started to increase during the rising phase (from MJD 59216) and has
continued to increase and reached a maximum value of $\sim $6~near the
peak of the outburst (MJD 59256), after that the HR started to decrease.
We have also studied the hardness intensity diagram using \textit{Swift}/BAT
and \textit{MAXI} flux. Figure~\ref{fig:HID} shows the hardness intensity diagram
(HID) for 2S 1417--624 during the outburst. The HID shows that there is
a sudden turn towards the left above the critical luminosity. The low luminosity
states are represented by the horizontal branch (HB) and the high luminosity
states are shown by the diagonal branch (DB). This sudden turn above the
critical luminosity implies a state transition from subcritical to supercritical
for this source.

\begin{figure}[t]
\centering{
\includegraphics[width=8cm]{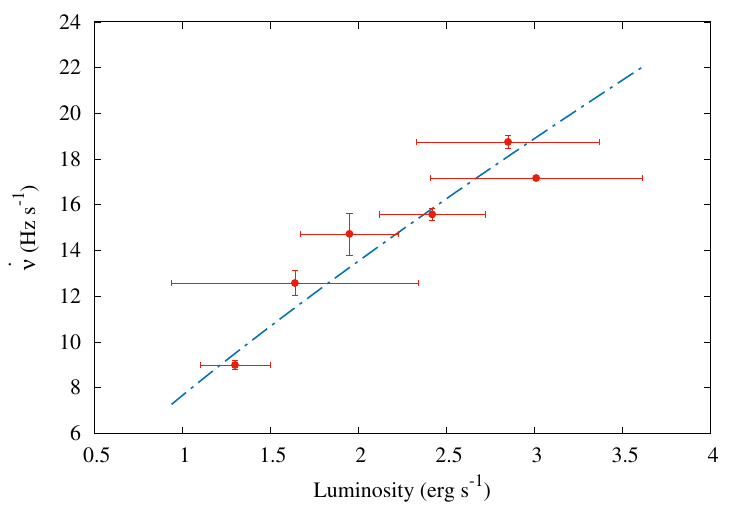} 
	\caption{Variation of spin change rate (in unit of $10^{-12} \text{Hz}\,\text{s}^{-1}$) with luminosity (in unit of $10^{37}$~erg\,s$^{-1}$). The dotted blue line represents the best power-law fit of data points that gives a power-law index of 0.82$\pm $0.11.}
\label{fig:GL}}
\end{figure}

Figure~\ref{fig:GL} shows the variation of spin-up rate with luminosity.
A power law is used to fit the $\dot{\nu}$ and luminosity. This shows a
positive correlation between the spin-up rate and the luminosity. Spin
frequency derivatives vary between $\sim $(0.8--1.8)$\times 10^{-11}\text{ Hz}\,\text{s}^{-1}$, which is estimated from spin frequency evolution history as
provided by \textit{Fermi}/GBM. The luminosity is varied between
$\sim $(1.0--3.5)$\times 10^{37}$~erg\,s$^{-1}$, which is estimated from
the \textit{Swift}/BAT count rate using a multiplying factor.

\begin{table*}
\centering
	\caption{Spectral fitting parameters of 2S 1417--624 for best-fit model
using \textit{NICER} (0.8--12~keV) observations:}
	
\label{tab:Spec}
\scalebox{0.55}{
	\begin{tabular}{cccccccccccc} \\
\hline
Obs. ID & Date & Count & $N_{H}$ & Photon index & E$_{cut}$ &  E$_{fold}$  &$kT_{bb}$ & Line & Equivalent & Reduced $\chi{^2}$ & Unabsorbed flux  \\
       &      & rate  &         &  ($\Gamma$)    &  & &         & energy & width     & (d.o.f)    & (10$^{-10}$) \\
       &(MJD) &(Count s$^{-1}$)  &( 10$^{22}$, cm$^{-2}$)&     & (keV)      & (keV)       & (keV)     & (keV) &  (keV)& & (erg cm$^{-2}$ s$^{-1}$) \\ 
\hline
3200130112 & 59241.50   & 39.85$\pm$0.08 & 1.20$^{+0.08}_{-0.07}$ & 0.41$^{+0.03}_{-0.03}$ & 6.39$^{+0.35}_{-0.35}$ & 11.4$^{+1.7}_{-1.3}$ & 0.255$^{+0.05}_{-0.03}$& 6.39$^{+0.04}_{-0.05}$ & 0.02$^{+0.11}_{-0.11}$ & 1.07 (935) & 6.539$\pm$0.004    \\
1200130165 &58308.25 &40.23$\pm$0.15 & 1.19$^{+0.13}_{-0.10}$ & 0.45$^{+0.03}_{-0.03}$ & 5.83$^{+0.44}_{-0.37}$ &18.75$^{+4.56}_{-3.13}$ &0.3$^{+0.12}_{-0.09}$ &6.42$^{+0.08}_{-0.07}$ & 0.07$^{+0.11}_{-0.08}$& 1.03 (754) & 7.398$\pm$0.006   \\
1200130155 &58296.28 &52.06$\pm$0.18 & 1.14$^{+0.09}_{-0.07}$ & 0.51$^{+0.02}_{-0.02}$ & 5.75$^{+0.5}_{-0.3}$ &18.04$^{+7.9}_{-4.8}$ &0.39$\pm$0.11 &6.46$^{+0.07}_{-0.06}$ & 0.12$^{+0.11}_{-0.06}$& 1.02 (780)& 9.46$\pm$0.90   \\
1200130166 &58310.32 &36.91$\pm$0.20 & 1.15$^{+0.16}_{-0.11}$ & 0.40$^{+0.04}_{-0.04}$ & 4.95$^{+0.86}_{-0.37}$ &14.07$^{+10.5}_{-5.6}$ &0.29$^{+0.17}_{-0.23}$ &6.4$^{+0.13}_{-0.07}$ & 0.04$^{+1.97}_{-0.04}$& 1.04 (607)& 6.467$\pm$0.009   \\
1200130168 & 58312.44&33.26$\pm$0.17 & 1.02$^{+0.27}_{-0.10}$ & 0.43$^{+0.05}_{-0.05}$ & 5.80$^{+1.1}_{-1.7}$ &15.53$^{+10.4}_{-2.0}$ &0.15$^{+0.3}_{-0.2}$ &6.39$^{+0.03}_{-0.04}$ & 0.0002$^{+0.08}_{-0.0001}$& 0.97 (635)& 5.935$\pm$0.009  \\
1200130169 &58317.70 &26.27$\pm$0.20 & 1.43$^{+0.32}_{-0.33}$ & 0.46$^{+0.05}_{-0.05}$ & 5.93$^{+1.1}_{-0.96}$ &10.95$^{+8.4}_{-4.1}$ &0.20$^{+0.19}_{-0.2}$ &6.39$^{+1.2}_{-0.9}$ & 0.07$^{+1.2}_{-0.07}$& 1.01 (530)& 4.560$\pm$0.09  \\
1200130104 &58214.77 &36.69$\pm$0.19 & 1.38$^{+0.17}_{-0.18}$ & 0.30$^{+0.05}_{-0.05}$ & 7.1$^{+1.6}_{-1.4}$ &12.7$^{+9.9}_{-6.5}$ &0.23$^{+0.08}_{-0.04}$ &6.43$^{+0.21}_{-0.17}$ & 0.11$^{+0.3}_{-0.1}$& 0.97 (630)& 6.90$\pm$0.01  \\
1200130143 &58215.45 &62.65$\pm$0.24 & 1.13$^{+0.05}_{-0.05}$ & 0.45$^{+0.11}_{-0.11}$ & 7.2$^{+1.5}_{-1.5}$ &12.5$^{+3.5}_{-3.5}$ &0.42$^{+0.09}_{-0.09}$ &6.39$^{+0.05}_{-0.05}$ & 0.20$^{+0.05}_{-0.05}$& 0.97 (754)& 11.63$\pm$0.90  \\
1200130144 & 58274.33 &60.10$\pm$0.20 & 1.14$^{+0.05}_{-0.05}$ & 0.52$^{+0.05}_{-0.05}$ & 9.67$^{+0.73}_{-0.73}$ &14.0$^{+3.03}_{-3.03}$ &0.40$^{+0.04}_{-0.04}$ &6.40$^{+0.05}_{-0.05}$ & 0.05$^{+0.04}_{-0.04}$& 0.97 (807)& 11.55$\pm$0.90  \\
1200130175 &58326.12 &22.64$\pm$0.13 & 1.13$^{+0.05}_{-0.05}$ & 0.45$^{+0.05}_{-0.05}$ &6.32$^{+0.45}_{-0.45}$ &11.60$^{+2.7}_{-2.7}$ &0.15$^{+0.04}_{-0.04}$ &6.42$^{+0.03}_{-0.03}$ & 0.002$^{+0.05}_{-0.05}$& 1.03 (596)& 3.95$\pm$0.96  \\
1200130177 &58328.82 &22.73$\pm$0.02 & 1.08$^{+0.05}_{-0.05}$ & 0.40$^{+0.07}_{-0.07}$ & 6.37$^{+0.4}_{-0.4}$ &7.05$^{+1.3}_{-1.3}$ &0.11$^{+0.05}_{-0.05}$ &6.42$^{+0.05}_{-0.05}$ & 0.02$^{+0.09}_{-0.09}$& 1.04 (496)& 3.90$\pm$0.90  \\
\hline
	\end{tabular} }
	\begin{tablenotes}
		\small
	\item  { Model }: {phabs$\times$(powerlaw$\times$highEcut+bbodyrad+gaussian)},  
	$N_{H}$ : hydrogen column density, 
	$\Gamma$: power-law photon index, $kT_{bb}$ : blackbody temperature,
 All the flux (unabsorbed flux) values  quoted in the paper are calculated  by using the {\tt cflux} convolution model. All of the reported errors were obtained using the  {\tt err} tool from {\tt XSPEC}. Uncertainties are given for a 90\% confidence interval.
	\end{tablenotes}
\end{table*}

\subsection{Energy spectrum}
The spectra were produced using the \textit{NICER} (0.8--12~keV) data. We
have excluded the spectrum above 12~keV and below 0.8~keV due to the poor
source count rate statistics in these ranges. The \textit{NICER} spectra were
fitted using \texttt{XSPEC} v12.11.0 and model parameters were varied independently.
We tested different simple single-component models like high energy \texttt{cut-off
power-law}, \texttt{bbody}, and \texttt{compTT} as well as combinations of models
like \texttt{power-law+bbody}, \texttt{diskbb+bknpower}, and\break \texttt{po}$\times $\texttt{highEcut+bobdyrad} to fit the source spectra. The spectra were well fitted with a blackbody emission and a power-law component with a
high-energy cut-off and an iron emission line at 6.4~keV, also modeled
using a Gaussian. The blackbody has been introduced along with a simple
power-law to model the soft excess component \citep{Hi04ApJ}. To find the
effect of absorption by hydrogen, all model components were multiplied
by a photo-electric absorption model.

We have studied the energy spectra using \textit{NICER}/XTI data and compared
the variation of spectral parameters near the same flux levels with the
earlier outburst in 2018. Earlier, the spectra of the source were modeled
using an absorbed power-law with a high energy cut-off and an iron emission
line near 6.4~keV \citep{Gu19}. We have applied an absorbed power-law continuum
model, and an additional blackbody emission has been introduced, which
improves the fit statistics. This model is good enough to describe the
spectral continuum at lower flux limits, which becomes more complex at
higher flux levels, as observed during the giant outburst of 2018
\citep{Gu19}. The additional emission near 6.4~keV is modeled using a
Gaussian component, which provides a reduced $\chi ^{2}$ value of
$\sim $1.0.

\begin{figure}
\centering{\includegraphics[width=5.5cm,angle=270]{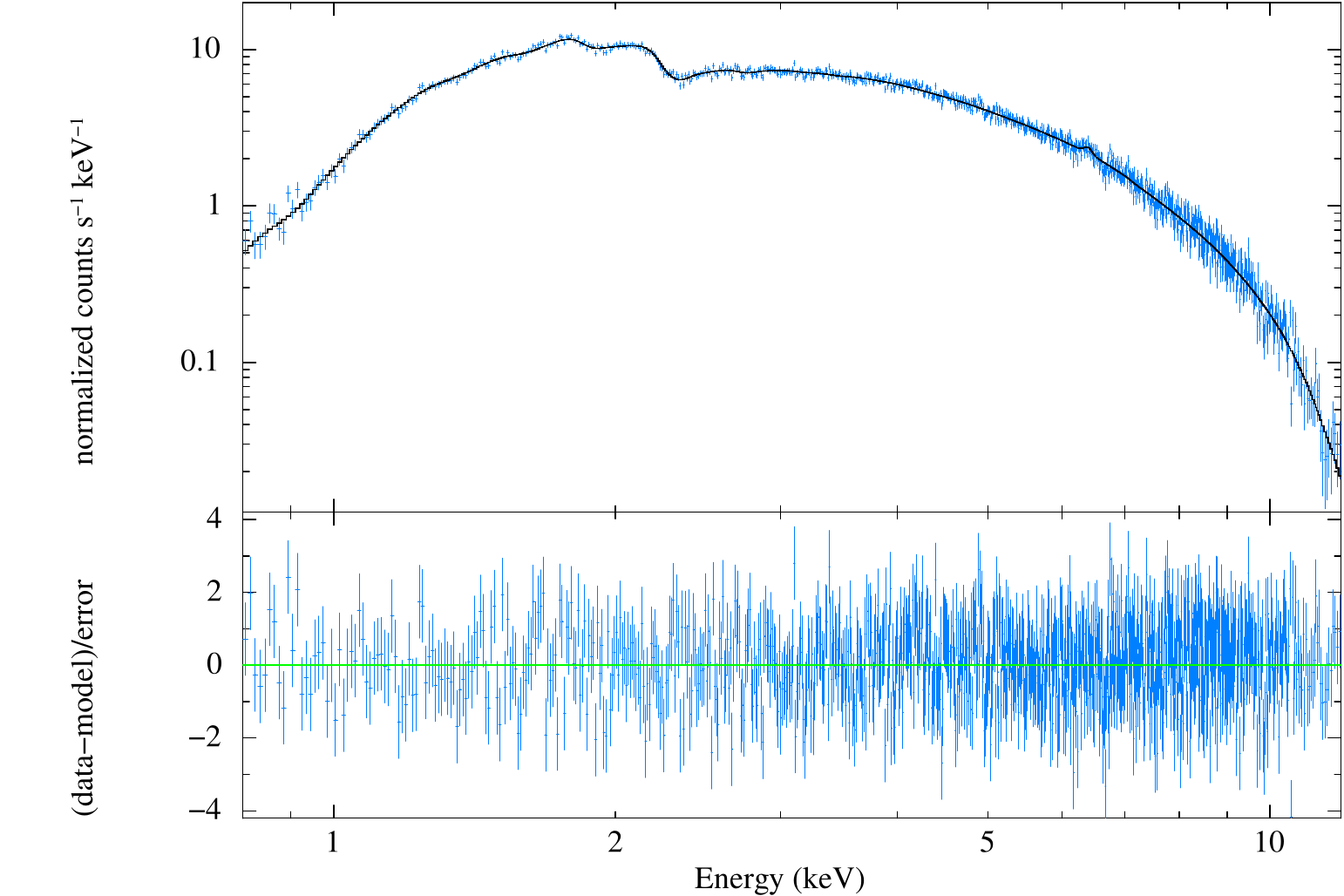}
	\caption{The energy spectrum of \textit{NICER} (0.8--12~keV) with the best-fitted model \textit{phabs}$\times $(\textit{powerlaw}$\times $\textit{highEcut+bbodyrad+gaussian}) during 2021 outburst. The residual is shown in the bottom panel.}
	\label{fig:spectrum}
	}
\end{figure}

The energy spectrum with the best-fitted models during the 2021 outburst
is shown in Fig.~\ref{fig:spectrum} where the bottom panel of Fig.~\ref{fig:spectrum} shows the residuals. The energy spectrum of the X-ray
pulsar can be well fitted with a high-energy cut-off power-law (\texttt{power-law}$
\times $\texttt{highEcut} in \texttt{XSPEC}) and a blackbody emission component
\texttt{bbodyrad} in \texttt{XSPEC}) along with a photoelectric absorption
(\texttt{phabs} in \texttt{XSPEC}). The \textit{NICER} spectrum in the energy range
0.8--12~keV is well described with blackbody emission with temperature
($kT_{bb}$) $\sim $ 0.255~keV and hydrogen column density $\sim 1.2\times 10^{22}~\text{cm}^{-2}$.

The unabsorbed X-ray flux in the 0.8--12~keV energy range is $\sim 6.5
\times 10^{-10}$~ergs\,cm$^{-2}$\,s$^{-1}$ using \textit{NICER} observation
during the 2021 outburst. Table~\ref{tab:Spec} summarizes the values of
different spectral parameters for the best-fitted model. The \texttt{cflux}
convolution model was used to calculate all of the flux (unabsorbed) values
in the paper. All of the reported errors were obtained using the err tool
from \texttt{XSPEC}. Uncertainties are given for a 90\% confidence interval.
We have also looked at the variation of spectral parameters with X-ray
flux from different \textit{NICER} observations in 2018 and 2021.

\begin{figure*}
\centering{
 \includegraphics[width=8.5cm]{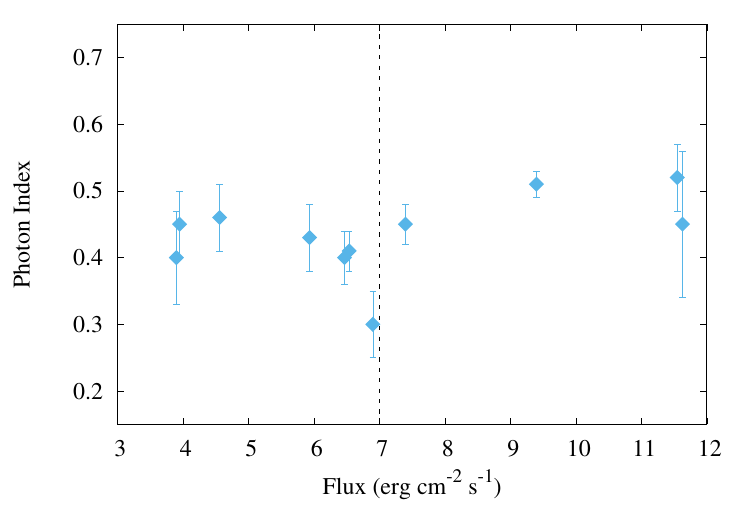}
 \includegraphics[width=8.5cm]{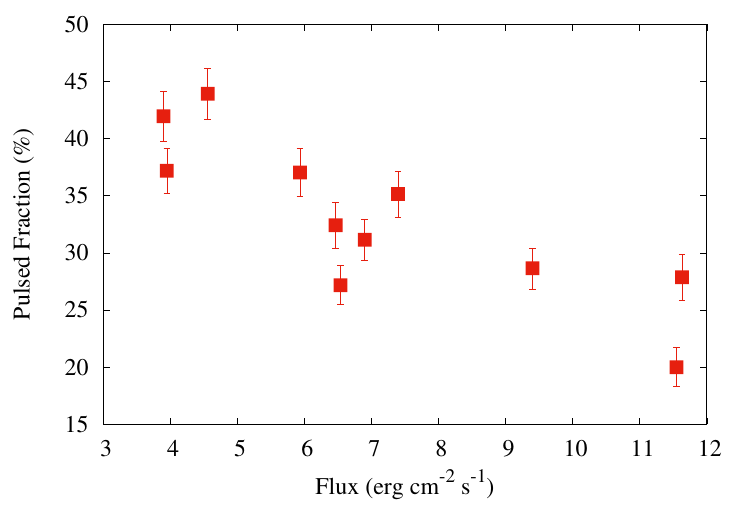}
\caption{The left-hand side image shows the variation of photon index with
X-ray flux using \textit{NICER} observations. The flux is estimated for the
range 0.8--12~keV in unit of $10^{-10}$~erg\,cm$^{-2}$\,s$^{-1}$. The vertical
dotted lines indicate the critical luminosity above which the correlation is changed. The figure shows that below the flux level of $\sim 7\times 10^{-10}$~erg\,cm$^{-2}$\,s$^{-1}$ the photon index shows an anti-correlation with \textit{NICER} flux and near the critical value, the photon index shows a trend to increase slightly. The point of a flux value of $6.539 \times 10^{-10}$~erg\,cm$^{-2}$\,s$^{-1}$ is during the 2021 outburst and the other points are during the 2018 outburst. The right-hand side image shows the variation of the pulsed fraction with \textit{NICER} flux.} 
\label{fig:PIflux}}	
\end{figure*}

The photon index decreased as the flux increased and shows an anti-correlation
below the flux level of $\sim 7\times 10^{-10}$~erg\,cm$^{-2}$\,s$^{-1}$,
which implies that at the brighter phase of the source the X-ray emission
was harder. Near the peak of the outburst during \textit{NICER} observation,
the photon index was $\sim $0.4. For the sake of comparison, we have used
a few earlier \textit{NICER} data points at nearly the same flux level. The
photon index showed a consistent value with the earlier observation at
the same flux level. Hydrogen column density and the blackbody temperature
also did not show any significant variation at the same flux level. The
radius of the emitting region of the blackbody is estimated using the normalization
constant of the model \texttt{bbodyrad} as norm $=$ $R_{km}^{2}/D_{10}^{2}$,
where $R_{km}$ is the source radius in km and $D_{10}$ is the distance
to the source in units of 10~kpc. The BB emission region is found to be
$\sim $12~km at a flux level of $\sim 6.54 \times 10^{-10}$~erg\,cm$^{-2}$\,s$^{-1}$ using \textit{NICER} (0.8--12~keV) observation during 2021 outburst.
Earlier, during the 2018 outburst, the BB emission radius was estimated
to be $\sim $8~km using \textit{NICER} observations for a source distance
of $\sim $9.9~kpc using \textit{bbodyrad} model \citep{Se22} at a flux level
of $\sim 19.36\times 10^{-10}$~erg\,cm$^{-2}$\,s$^{-1}$ in the energy of
0.8--12~keV of \texttt{NICER}. The variation of photon index with flux is
shown in Fig.~\ref{fig:PIflux}, which shows that the photon index is anti-correlated
with the X-ray flux below the critical flux level ($\sim 7\times 10^{-10}$~erg\,cm$^{-2}$\,s$^{-1}$), which is shown by a vertical dotted line. Above
the critical flux, the correlation turns to a slightly positive trend,
which may indicate a state transition for this source.

\section{Discussion}
\label{dis}
We present the results of timing and spectral analysis of 2S 1417--624
using the \textit{NICER} data during the recent outburst in 2021. The timing
analysis reveals that the pulse profile shows multiple peaks and dips with
an energy-dependent nature, which is comparable with the previous results
at the same flux levels during the giant outbursts in 2009
\citep{Gu18} and 2018 \citep{Ji20,Gu19}. The energy dependence of the
pulse profile is studied to investigate the evolution of the individual
dips and peaks of the pulsar with different energies. The pulse profile
evolves significantly with energy, which is comparable at the same flux
level as the previous result \citep{Ji20}. During the 2021 outburst, the
spin-up rate is found to vary from $0.8\times 10^{-11}\text{ Hz}\,\text{s}^{-1}$ to
$1.8\times 10^{-11}\text{ Hz}\,\text{s}^{-1}$, which is comparable to the previous
outbursts of 2S 1417--624 \citep{Fi96a,Ji20}. The mass accretion rate
is estimated using the luminosity near the peak (during \textit{NICER} observation)
of the outburst as, $L = \eta \dot{M}c^{2}$ and $\dot{M}=1.3
\times 10^{17}\text{ g}\,\text{s}^{-1}$, which is estimated using the accretion efficiency
factor $\eta =0.2$. The variation of pulsed fraction with energy also
shows consistent values at comparable flux levels in the 2018 and 2021
outbursts. The value of the pulsed fraction in the 6--10~keV energy band
is not consistent probably due to the low count in this band. The pulsed
fraction showed a negative correlation with luminosity. We have used a
few earlier \textit{NICER} data along with recent data to investigate the
evolution of pulsed fractions at different flux levels.

Earlier, the X-ray pulsar 2S 1417--624 showed strong luminosity and energy
dependence during the 2018 giant outburst. During the present outburst,
two broad peaks and dips are observed in the pulse profile, and the pulse
profile evolves with energy. The double peak feature, with 0.5 separations
of each peak, indicates a simple beam function created from both of the
poles of the neutron star during the outburst \citep{Gu18}. During the
2018 giant outburst, the luminosity dependency of the pulse profile and
the complex shape of the pulse profiles were reported by \citet{Ji20}.
In the supercritical regime, an additional dip was observed, probably due
to the increased absorption at higher luminosity. It is also possible that
the hydrogen column density of the partial covering absorber ($n_{H2}$)
may increase due to the absorption of the materials around the neutron
star as in the case of Swift J0243.6+6124 \citep{Zh19}. At higher luminosity,
the beam patterns become more complex and mostly dominated by fan-beam
or a mix of pencil and fan-beam patterns. During the 2009 giant outburst,
\citet{Gu19} performed phase-resolved spectroscopy for this source and
concluded that except for the primary dip (phase 0.95-1.05) there was no
significant variation in the additional column density. Several pulsars
showed strong energy and luminosity dependence of the pulse profiles, like
1A 0535+262 \citep{Ma22}, EXO 2030+375 \citep{Ep17,Na13}, GX 304--1
\citep{Ja16}.

The luminosity during the \textit{NICER} observation of the 2021 outburst
was below the critical luminosity for this source, and at this phase of
the outburst, the pulse profile showed two broad peaks, which is consistent
with the previous result \citep{Ji20} and the emission geometry is expected
to be driven by a ``pencil beam'' pattern during this sub-critical accretion
regime. During the recent outburst, multiple dips-like features were observed
from 2S 1417--624. The typical value of the critical luminosity of accretion-powered
X-ray pulsars is estimated to be in the order of $10^{37}$~erg\,s$^{-1}$
\citep{Be12,Re13}. This critical value of luminosity is essential to understand
the transition between sub-critical and super-critical accretion regimes.
In sub-critical phases, the emission geometry is comparatively simple,
and the accretion flow is supposed to halt by Coulomb interaction close
to the neutron star surface. A single or double-peaked pulse profile is
produced, and emission geometry can be driven by a pencil beam-like pattern.
Earlier, in the 2018 outburst, the luminosity was higher and near the critical
luminosity, the beam geometry changed and multiple peaks were observed
in the pulse profile \citep{Gu18}. The emission geometry was changed from
a pencil beam to a mixture of pencil and fan beams.

We look at the hardness ratio (HR) using the ratio of the count rates from
\textit{Swift}/BAT (15--50~keV) and \textit{MAXI}/GSC (2--20~keV) to study the
evolution of the spectral states during the outburst. The HR shows variability
during the outburst and the HR varied between $\sim $0.1--6 for the time
MJD 59216--59326. We have also studied the hardness intensity diagram (HID) to
look for any changes in spectral shape. The HID shows a state change from
a subcritical to a supercritical accretion regime. In the HID for the pulsar
2S 1417--624, we noticed a change from HB to DB. Earlier different pulsars,
such as 4U 0115+63, EXO 2030+375, V 0332+53, KS 1947+300, and 1A 0535+262
showed a transition in their HID \citep{Re13, Ma22}. When the luminosity
reached a critical value, 2S1417--624 entered the DB by making a sudden
left turn in the HID. The hardness ratio (HR) started to drop above the
critical luminosity, and the peak of the outburst corresponded to the softest
state of DB.

Earlier, critical flux level was estimated to be $\sim 0.7\times 10^{-9}$~erg\,cm$^{-2}$\,s$^{-1}$ using \textit{NICER} (0.8--12~keV) for a source distance
of 9.9 kpc \citep{Se22}. During the \textit{NICER} observation of the 2021
outburst, the unabsorbed flux (0.8--12~keV) level was below the critical
level. Above this flux level, the source state transition from sub-critical
to super-critical accretion regime may occur, and in this regime, the radiation
pressure is high enough to halt the accretion flow at a certain height
above the pulsar. During this transition, the pulse profile morphology,
pulse fraction, and different spectral parameters also show significant
variation. The beaming pattern also seems to change from pencil-beam to
fan-beam or a mix of pencil and fan-beam \citep{Be12}.

Earlier, a correlation between spin-up rate and X-ray flux was observed
during outbursts for different transient systems, which was explained in
terms of accretion. For example, 2S 1417--624 \citep{Fi96a}, A 0535+26
\citep{Fi96b, Bi97}, EXO 2030+375 \citep{Pa89, Re96}, GRO J1744--28
\citep{Bi97}, and SAX J2103.5+4545 \citep{Ba02} showed correlation between
spin-up rate and X-ray flux. Figure~\ref{fig:GL} shows that the pulse frequency
derivatives of the X-ray pulsar 2S 1417--624 are correlated with luminosity.
Earlier, for 2S 1417--624, \citet{Fi96a} observed that $\dot{\nu}$ was
highly correlated with the pulsed flux. Both these parameters are supposed
to be driven by the mass accretion rate.

Based on the accreting torque model and the observed spin-up rate, we have
tried to find the magnetic dipole moment and the surface magnetic field
of 2S 1417--624. The spin-up rate and the luminosity are known to be correlated
in transient X-ray pulsars as \citep{Gh79b, Su17}:
%
\begin{equation}
\dot{\nu}_{12}=2.0 n \zeta ^{\frac{1}{2}} \mu _{30}^{\frac{2}{7}} R_{6}^{
\frac{6}{7}} M_{1.4}^{-\frac{3}{7}} I_{45}^{-1} L_{37}^{\frac{6}{7}}
\end{equation}
where $\dot{\nu}_{12}$, $\mu _{30}$, $R_{6}$, $M_{1.4}$, and
$I_{45}$ are the spin frequency derivative, magnetic dipole moment, radius,
mass and the moment of inertia of the neutron star given in the units of
$10^{-12}\text{ Hz}\,\text{s}^{-1}$, $10^{30}$~G\,cm$^{3}$, $10^{6}$~cm, 1.4~$M_{\odot}$, and
$10^{45}\text{ g}\,\text{cm}^{2}$ respectively. $L$ is the X-ray luminosity
in the unit of $10^{37}$~erg\,s$^{-1}$. According to the
\citet{Gh79a,Gh79b} model, under slow-rotator condition, $n\sim $1.39
and $\zeta \sim 0.52$. Therefore, equation (1) reduces to
\citep{Su17}
%
\begin{equation}
\dot{\nu}_{12}=k L_{37}^{\alpha}
\end{equation}
where $k = 2.0 \mu _{30}^{\frac{2}{7}}$ and $\alpha  =
\frac{6}{7}$. For nominal values of $R_{6} = M_{1.4} = I_{45} = 1$, measurements
of the $\dot{\nu}$ versus $L$ give a rough estimation of the magnetic dipole
moment of the pulsar. From the $L$ vs $\dot{\nu}$ plot, we have estimated
$k$ and $\alpha $ as 7.65$\pm $0.71 and 0.82$\pm $0.11 respectively from
the best fit result. The estimated value of $\alpha $ is close to the theoretical
value. Figure~\ref{fig:GL} shows the correlation between the spin-up rate
and luminosity and the solid line represents the best-fitted result. From
the best fit result, we may write the equation (2) as
%
\begin{equation}
\dot{\nu}_{12}=(7.65\pm 0.71) L_{37}^{0.82\pm 0.11}
\end{equation}
Now the magnetic dipole moment can be written in the form
%
\begin{equation}
2.0 \mu _{30}^{\frac{2}{7}} = 7.65 ; \mu _{30} \simeq 109
\end{equation}

The surface magnetic field can be estimated using the magnetic moment ($
\mu _{30}$) and radius ($R_{6}$) of the pulsar as
%
\begin{equation}
\mu _{30} = \frac{1}{2} B_{12}R_{6}^{3}\phi (x)
\end{equation}
$\phi (x)$ is the correlation factor, for typical NS, $\phi (x)$
$\sim $0.68, the magnetic field can be written as
%
\begin{equation}
B_{12} = 2 \times \frac{\mu _{30}}{0.68}
\end{equation}
for $\mu _{30}$ $\simeq $109, the magnetic field is estimated to be
$\simeq  3\times  10^{14}$~G.

The high value of the magnetic dipole moment leads to a higher value of
the magnetic field ($\sim 10^{14}$~G). Earlier, \citet{Ji20} also concluded
that the estimated magnetic field of this source was high during the 2018
giant outburst. If the source distance is taken as twice ($\sim $20 kpc
\citep{Ji20}) of the Gaia estimated distance, then the magnetic field strength
reduces to a typical value of the order $\sim 10^{12}$~G.
\citet{Se22} also concluded that the magnetic field of the source was very
high ($\sim 10^{14}$~G) like a magnetar during another study of the 2018
outburst, which is consistent with our results in the 2021 outburst. The
high magnetic field in 2S 1417--624 may originate from the limitation of
torque models, which do not allow closer distance \citep{Ma2020ApJ}. There
are several sources like XTE J1858+034, GRO J1008--57, GS 0834--430, IGR
J18179--1621, IGR J19294+1816, RX J0440.9+4431, \textit{MAXI} J1409--619,
and GRO J2058+42 for which considerable deviations from the GL model were
observed, even considering the Gaia measured distances
\citep{Ma2020ApJ}. During the 2009 outburst of 2S 1417--624, the critical
luminosity was estimated to be $\sim 1.33 \times 10^{37}$~erg\,s$^{-1}$
(3--30~keV flux, source distance of 11 kpc) by assuming a magnetic field
of $0.9\times 10^{12}$~G \citep{Ia04, Gu18}. \citet{Se22} found that the
critical flux during the 2018 outburst was $\sim 7 \times 10^{-10}$~erg\,cm$^{-2}$\,s$^{-1}$ above which spectral parameters showed significant evolution.
We have found that the critical flux (0.8--12~keV) during the 2021 outburst
to be $\sim 7 \times 10^{-10}$~erg\,cm$^{-2}$\,s$^{-1}$, which is consistent
with the previous outburst \citep{Se22}. We have also estimated the magnetic
field corresponding to the critical luminosity using equation 8, during
the 2021 outburst of the source. The magnetic field corresponding to the
critical luminosity $0.82\times 10^{37}$~erg\,s$^{-1}$ is estimated to be
$0.57 \times 10^{12}$~G for a source distance of 9.9 kpc \citep{Ba18}.
For a typical neutron star, the critical luminosity and magnetic field
are associated as \citep{Be12},
%
\begin{equation}
L_{\mathrm{critical}} = 1.5 \times 10^{37}\left (\frac{B}{10^{12} G}
\right )^{\frac{16}{15}}~\text{erg}\,\text{s}^{-1}
\end{equation}
Using the torque luminosity model, the magnetic field was estimated to
be $\sim 7\times 10^{12}$~G and the distance was estimated to be
$\sim $20~kpc \citep{Ji20}.

Earlier, in 2013, Chandra observed 2S 1417--624 during the quiescent phase.
The pulsar spectrum was characterized by either a power-law or a blackbody
model with a high temperature of $\sim $1.5~keV \citep{Ts17}. The spectrum
of the source during the 2021 outburst was well explained by a composite
model of power-law and blackbody components. During the 2018 giant outburst,
the source spectrum was well explained with the cut-off power-law continuum
model and a blackbody component with the interstellar absorption
\citep{Gu19}. The energy spectrum (\textit{NICER}/XTI) of the source during
the recent outburst in 2021 is also well described with a similar type
of model as observed earlier. We have compared the spectral properties
of the X-ray pulsar with the 2018 giant outburst at a comparable flux level.
Earlier, the photon index showed an anti-correlation with flux below
$\sim 0.7 \times $ 10$^{-9}$~erg\,cm$^{-2}$\,s$^{-1}$. Such an anti-correlation
was also observed in both the 1999 \citep{Ia04} and 2009 outbursts
\citep{Gu18}. During the recent outburst of 2021, the \textit{NICER} flux
was below the critical value ($\sim 0.7 \times 10^{-9}$~erg\,cm$^{-2}$\,s$^{-1}$), and the photon index showed an anti-correlation with X-ray flux
below critical luminosity. During the 2018 outburst, \citet{Se22} also
found an anti-correlation between the photon index and flux below the critical
flux value, which turns into a slightly positive correlation above the
critical value of flux. We have compared the \textit{NICER} spectra during
the 2018 and 2021 outbursts at comparable flux levels, and the results
show consistent values of spectral parameters and the photon index decreases
with the increase of X-ray flux and the spectrum gets harder.

We have found the radius of the emitting region of the pulsar using the
fitting parameters. The normalization constant of the \texttt{bbodyrad} model
gives the size of the emission region for a known distance to the source.
From the results of spectral fitting during the 2021 \textit{NICER} observation,
the radius of the emitting region is $\sim $12~km for a distance of
$\sim $9.9 kpc.

Earlier, a significant change in the correlation of the $L$~--~$\Gamma $ diagram was seen in different sources near the critical luminosity.
The transition from a negative to positive correlation was seen in the
$L$~--$~\Gamma $ diagram as luminosity increases above the critical luminosity
\citep{Re13}. In the subcritical regime, a negative correlation was reported
for the sources like 1A 1118--612, GRO J1008--57, XTE J0658--073, and a
transition in the correlation of $L$~--$~\Gamma $ was observed for the
sources 1A 0535+262 \citep {Ma22}, 4U 0115+63, EXO 2030+375
\citep{Ep17,Ja21}, 2S 1417--624 \citep{Se22}, and KS 1947+300
\citep{Re13}. In the subcritical accretion regime, the negative correlation
implied the hardening of the power-law continuum with flux. In the supercritical
accretion regime, the positive correlation implied the softening of the
power-law continuum with flux.
\section{Conclusions}
\label{con}
We have summarized the results of the timing and spectral analysis of the
X-ray pulsar 2S 1417--624 during the outburst in 2021. The spin-up rate
varied between $\sim 0.8\hbox{--}1.8\times 10^{-11}\text{ Hz}\,\text{s}^{-1}$ during the outburst. A positive correlation is observed between the spin-up rate and
luminosity during the outburst. The torque-luminosity model gives a surface
magnetic field of the pulsar of $\sim 10^{14}$~G. The higher magnetic
field may arise due to the closer distance as given by Gaia. The pulse
profile showed multiple peaks and dips, which evolved with energy. The
energy spectrum of the source was well described with a composite model
consisting of a power-law with higher cut-off energy and a thermal blackbody
component. The radius of the emitting region of the pulsar is estimated
to be $\sim $12~km. The photon index showed an anti-correlation with X-ray
flux below the flux level of $\sim 7.0\times 10^{-10}$~erg\,cm$^{-2}$\,s$^{-1}$. During the outburst, the source state evolved from a subcritical
to a supercritical regime. The HID supported the state transition. A transition
from the horizontal to the diagonal branch is observed from the HID. As
flux increased, the spectrum became harder in the horizontal branch and
softer in the diagonal branch.

\section*{Acknowledgements}
We thank the anonymous reviewer for useful suggestions, which helped to
improve the manuscript significantly. This research has made use of the
\textit{MAXI} data provided by RIKEN, JAXA, and the \textit{MAXI} team. We acknowledge the use of public data from the \textit{NICER}, and \textit{Fermi} data archives.

\section*{Data Availability}
The data underlying this article are publicly available in the High Energy Astrophysics Science Archive Research Center (HEASARC) at \\
\url{https://heasarc.gsfc.nasa.gov/db-perl/W3Browse/w3browse.pl}.

\section*{Statements \& Declarations}
\subsection*{Funding}
The authors declare that no funds, grants, or other support were received during the preparation of this manuscript.

\subsection*{Competing Interests}
The authors have no relevant financial or non-financial interests to disclose.

\subsection*{Author Contributions}
All authors contributed to the study's conception and design. Data analysis was performed by Manoj Mandal and Sabyasachi Pal. The first draft of the manuscript was jointly written by both authors. All authors read and approved the final manuscript.


\makeatletter
\let\clear@thebibliography@page=\relax
\makeatother


\end{document}